\documentclass[twocolumn,showpacs,preprintnumbers,amsmath,amssymb]{revtex4}
\usepackage{amssymb}
\usepackage[dvips]{graphicx}
\begin{document}
\title{Hall effect and resistivity  in underdoped cuprates}
\author{A.S.~Alexandrov$^{1}$, V.N.~Zavaritsky$^{1,2}$, S.~Dzhumanov$^{1,3}$
}
\address{
$^1$Department of Physics, Loughborough University,Loughborough LE11\,3TU, UK\\
$^{2}$P.L.Kapitza Institute for Physical Problems, 2 Kosygina Str., 117973 Moscow, Russia\\
$^{3}$ Institute of Nuclear Physics,  702132 Tashkent, Uzbekistan}

\begin{abstract}
%\large
The behaviour of the Hall ratio $R_{H}(T)$ as a function of temperature is one of the
most intriguing normal state properties of cuprate
superconductors. One feature of all the data is a maximum of $R_{H}(T)$
 in the normal state that broadens and shifts to temperatures well above $T_c$
 with decreasing doping.
We show that a model of preformed pairs-bipolarons
provides a selfconsistent quantitative description of  $R_{H}(T)$
together with    in-plane  resistivity and 
uniform magnetic susceptibility for a wide range of doping. 
\end{abstract}

\pacs{PACS:  74.72.-h, 74.20.Mn, 74.20.Rp, 74.25.Dw}
\vskip2pc]

\maketitle
%\Large
%\narrowtext

\bigskip

The theory of high temperature superconductivity  remains the biggest
challenge in condensed matter physics today. One way of thinking is that
this phenomenon is of purely electronic origin and
phonons are irrelevant \cite{PWA,KIVE,ZHAN,PINE}. Other authors
 (see, for example \cite{SANF,emi,dev,tru,kab,dzh}) explore an
 alternative view, namely that the extension of the BCS theory towards
 the strong interaction between electrons and ion vibrations describes
 the phenomenon. 
High temperature superconductivity could exist in the crossover
region of the electron-phonon interaction strength from the BCS to
bipolaronic superconductivity as was argued before the discovery \cite
{ale0}. In the strong coupling regime, $\lambda \gtrsim 1,$ pairing takes
place in real space (i.e. {\it individual pairing) } due to a polaron
collapse of the Fermi energy \cite{aleran}. At first sight, polaronic
carriers have a mass too large to be mobile; however it has been shown that
the inclusion of the on-site Coulomb repulsion leads to the favoured binding
of intersite oxygen holes \cite{ALEXAND,CATL}. The intersite bipolarons can
then tunnel with an effective mass of about 10 electron masses \cite{ALEXAND,CATL,alekor,tru}. 

The possibility of real-space pairing, as opposed to the Cooper pairing, has
been the subject of much discussion. Experimental \cite
{ZHAO,mic,ita,TIM,ega,LANZ} and theoretical \cite{phil,MOTT,all,gor}
evidence for an exceptionally strong electron-phonon interaction in all
novel superconductors is now so overwhelming that even some advocates of
non-phononic mechanisms \cite{kiv} accept this fact. Nevertheless, the
same authors \cite{KIVE,kiv} dismiss any real-space pairing, suggesting a 
{\it collective} pairing (i.e the{\it \ Cooper }pairs in the momentum space)
at some temperature $T^{\ast }>T_{c}$ but without phase coherence.  The existence of
noncoherent Cooper pairs might be a plausible idea for the crossover region
of the coupling strength, as was
proposed still earlier by Dzhumanov \cite{dzh}. However, apart from this, 
Refs.\cite{KIVE,kiv} argue that the phase coherence and
superconducting critical temperature $T_{c}$ are determined by the
superfluid density, which is proportional to doping $x$, rather than to the density of normal state
holes, which is ($1+x)$ in their scenario. On the
experimental side, the scenario  is not compatible with a great number
of thermodynamic, magnetic, and kinetic measurements, which show that only
holes {\it doped} into a parent insulator are carriers {\it in both} the
normal and superconducting state of {\it underdoped} cuprates. On theoretical
grounds, this preformed {\it Cooper}-pair (or phase-fluctuation) scenario
contradicts a theorem \cite{leg}, which proves that the number of
supercarriers (at $T=0$) and normal-state carriers should be the same in any
{\it clean} translation-invariant superfluid. A periodic crystal-field
potential does not affect  this conclusion because the
coherence length is larger than the lattice constant in cuprates. 
Objections against real-space pairing also contradict a
parameter-free estimate of the renormalised Fermi energy $\epsilon _{F}$\cite
{aleF}, that yields $\epsilon _{F}$ less than the normal state charge
pseudogap $\Delta /2$ in underdoped cuprates. The condition for real-space
pairing, $\epsilon _{F}\lesssim \pi \Delta ,$ is well satisfied if one
admits that the bipolaron binding energy, $\Delta$, is twice the pseudogap.

Bipolarons in cuprates could be formed by the Fr\"{o}hlich interaction of
holes with optical phonons \cite{ALEXAND}, and by molecular phonons (Jahn-Teller interactions \cite{kab}) because of molecular-like crystal
structure of these materials. Mott and Alexandrov proposed a
simple model \cite{MOTT} of the cuprates based on bipolarons. In this model,
all the holes (polarons) are bound into small intersite bipolarons at any
temperature. Above $T_{c}$ this Bose gas is non-degenerate and below $T_{c}$
phase coherence of the preformed bosons sets in, followed by
superfluidity of the charged carriers. Of course, there are also thermally
excited single polarons in the model. There is much evidence for the
crossover regime at $T^{\ast }\simeq \Delta/2$ and normal state charge and spin pseudogaps
in the cuprates \cite{mih}. Many experimental observations have been
satisfactorily explained using this particular approach including the
in-plane \cite{BRAT,in} and out-of-plane resistivity \cite{AKM,out}, magnetic
susceptibility \cite{AKM,MULLE}, tunneling spectroscopy \cite{NDROV}, isotope
effect \cite{ALEXA,ZHAO}, upper critical field and specific heat anomaly \cite
{ZAV}. ARPES measurements indicate the presence of a pseudogap as well. They
also indicate an angular dependent narrow peak and a featureless background.
In the polaronic model, the ARPES spectrum can be numerically explained if
one considers a charge transfer Mott-insulator and the single polaron
spectral function \cite{DENT}.

Like many other properties, the Hall ratio in high-$T_{c}$ cuprates shows a
non-Fermi-liquid behaviour \cite{tak,bat,car}. A Fermi-liquid approach
 may describe $R_{H}(T)$ for $T\gg T^*$ only if vertex corrections are included. 
However, the advocates of this approach \cite{kon} admit
that it is inappropriate for $T<T^{\ast }.$ On the other hand, the bipolaron
model described an enhanced magnitude, doping and $1/T$ temperature
dependence of the Hall ratio \cite{BRAT,ALEXAND}, but the maximum of $%
R_{H}(T) $ in underdoped cuprates well above $T_{c}$ \cite{tak,bat,car} was
not addressed. Also, a nonlinear temperature dependence of the $in$-plane
resistivity below $T^{\ast }$ remains one of the unsolved problems. In this
paper, we give an explanation of these long-standing problems from the
standpoint of the bipolaron model.

Thermally excited phonons and (bi)polarons are well decoupled in the
strong-coupling regime of electron-phonon interaction \cite{SANF}, so
that the conventional Boltzmann kinetics for renormalised carries is
applied. Here we use a `minimum' bipolaron model, which includes a singlet
bipolaron band and a spin 1/2 polaron band separated by $T^{\ast }$, and 
the $\tau -$approximation \cite{ANSE} in an electric{\rm \ }field ${\bf E}=-\vec{{\nabla}}\phi$ and in a weak magnetic field ${\bf B\perp }$ ${\bf E.}$
The bipolaron and single-polaron non-equilibrium distributions are found as 
\begin{equation}
f({\bf k})=f_{0}(E)+\tau \frac{\partial f_{0}}{\partial E}{\bf v}\cdot
\left\{ {\bf F}+\Theta {\bf n}\times {\bf F}\right\} ,  \label{4}
\end{equation}
where ${\bf v=}\partial E/\partial {\bf k,}$ ${\bf F}={\vec{\bf\nabla}}(\mu
-2e\phi )$ and $f_{0}(E)=[y^{-1}{\exp (E/T)-1]}^{-1}$ for bipolarons with
the energy $E=k^{2}/(2m_{b})$, and ${\bf F}={\vec{\bf\nabla}}(\mu /2-e\phi
) $ and $f_{0}(E)=\{y^{-1/2}{\exp [(E+T^{\ast })/T]+1\}}^{-1}$,
$E=k^{2}/(2m_{p})$ for thermally excited polarons. Here $m_{b}\approx
2m_p$ and $m_p$ are the
bipolaron and polaron masses of $quasitwo$-dimensional carriers, $y=\exp
(\mu /T),$ $\mu $ is the chemical potential, $\hbar =c=k_{B}=1$, and ${\bf %
n=B/}B$ is a unit vector in the direction of the magnetic field. Eq.(1) is
used to calculate the electrical resistivity and the Hall ratio as 
\begin{eqnarray}
\rho &=&\frac{m_{b}}{4e^{2}\tau _{b}n_{b}(1+An_{p}/n_{b})}, \\
R_{H} &=&\frac{1+2A^{2}n_{p}/n_{b}}{2en_{b}(1+An_{p}/n_{b})^{2}},
\end{eqnarray}
where $A=\tau _{p}m_{b}/(4\tau _{b}m_{p})$. The atomic densities of carriers
are found as
\begin{eqnarray}
n_{b}=\frac{m_{b}T}{2\pi }|\ln (1-y)|, \\
n_{p}=\frac{m_{p}T}{\pi }\ln \left[ 1+y^{1/2}\exp \left( -T^{\ast }/T\right) %
\right] .
\end{eqnarray}
and the chemical potential is determined by doping $x$ using 
$2n_{b}+n_{p}=x-n_{L}$,
where $n_{L}$ is the number of carriers localised by disorder. Here we take
the lattice constant $a=1$. Polarons are not degenerate. Their number  remains small compared with twice
the number of bipolarons, $n_p/(2n_b)<0.2$, in the relevant temperature range
$T\lessapprox T^{\ast }$, so that
\begin{equation}
 y\approx 1-\exp(-T_0/T),
\end{equation}
where $T_0=\pi (x-n_L)/m_b \approx T_c$ is about the superconducting critical
temperature of the (quasi)two-dimensional Bose gas \cite{SANF}. Because of this reason, experimental $T_c$ was taken as  $T_0$ for our fits.

 Then using Eqs.(2,3) we obtain
\begin{equation}
R_{H}(T)=R_{H0}\frac{1+2A^2y^{1/2}(T/T_{c})\exp \left( -T^{\ast
      }/T\right) }{
[1+A(T/T_{c})y^{1/2}\exp \left( -T^{\ast }/T\right) ]^{2}},
\end{equation}
where $R_{H0}=[e(x-n_{L})]^{-1}$.  In the following, we assume that the number
of localised carriers depends only weakly on temperature in underdoped
cuprates because their average  ionisation energy is sufficiently large, so that 
$R_{H0}$ is temperature independent if $T\lessapprox T^{\ast }$.
As proposed in Ref.\cite{BRAT} the scattering rate ($\propto T^2$) is due to inelastic
 collisions of itinerent carriers with those localised by
 disorder. Here we also take into account the scattering off optical
 phonons \cite{ANSE}, so that 
$\tau^{-1}=aT^{2}+b\exp{(-\omega /T)}$,
if the temperature is low compared with the characteristic phonon
energy $\omega$. The relaxation times of each type of carriers scales with
their charge $e^\ast$ and mass as $\tau_{p,b} \propto m_{p,b}^{-3/2}(e^\ast)^{-2}$, so that $A=(m_b/m_p)^{5/2} \approx 6$ in the Born approximation for any short-range scattering potential. As a result we obtain the in-plane resistivity as 
\begin{equation}
\rho (T)=\rho _{0}\frac{(T/T_{1})^{2}+\exp \left( -\omega /T\right) }{[1+A(T/T_{c})y^{1/2}\exp \left( -T^{\ast }/T\right) ]},
\end{equation}
where $\rho _{0}=bm_{b}/[2e^{2}(x-n_L)]$ and $T_{1}=(b/a)^{1/2}$ are
temperature independent. Finally, we easily obtain the uniform magnetic
susceptibility due to nondegenerate spin 1/2 polarons as \cite{AKM} 
\begin{equation}
\chi (T)=By^{1/2}\exp \left( -T^{\ast }/T\right) +\chi _{0},
\end{equation}
where $B=(\mu _{B}^{2}m_{p}/\pi )$, and $\chi_{0} $ is the magnetic susceptibility of the parent Mott insulator.

The present model fits the Hall ratio, $R_H(T)$, the in-plane resistivity, $\rho(T)$, and the
magnetic susceptibility $\chi(T)$ of underdoped $YBa_{2}Cu_{3}O_{7-\delta }$ with a
selfconsistent set of parameters (see Fig. 1-2 and the Table). The ratio of polaron and
bipolaron mobilities $A=7$ used in all fits is close to the above
estimate,  and $\chi _{0}\approx 1.5\times 10^{-4}emu/mole$ is very close
to the susceptibility of a slightly doped insulator 
\cite{coop}.
The comprehensive analysis by
Mihailovic et al. \cite{mih} yileds $T^{\ast }$ in the range from $200K$ to $1000K$ depending on doping, and $\omega $ should be about $500K$ or so
from the optical data by Timusk et al. \cite{TIM}.

\begin{figure}
\begin{center}
\includegraphics[angle=-0,width=0.47\textwidth]{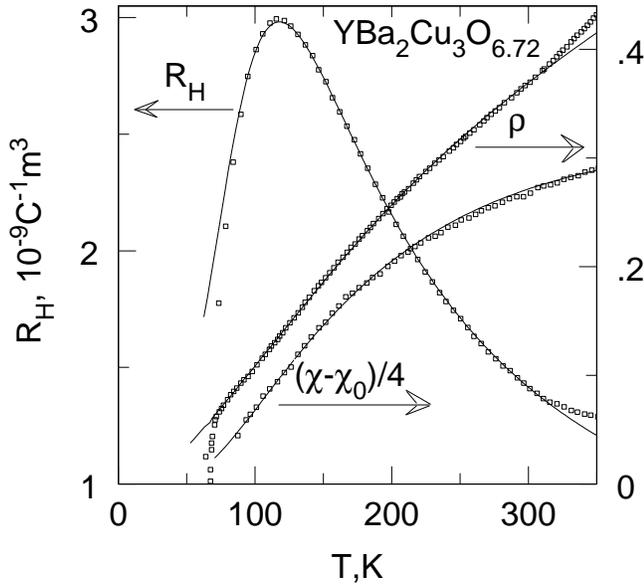}
\vskip -0.5mm
\caption{The Hall ratio and resistivity ($\rho, m\Omega cm$) of underdoped $YBa_{2}Cu_{3}O_{6.72}$ fitted by
the present theory; also shown the fit of rescaled susceptibility ($\chi, 10^{-4}emu/mole$) for similarly doped sample, $\delta=0.26$. (See Table for the parameters.)                                                
}
\end{center}
\end{figure}

\vspace{-5mm}\hskip -2mm
\begin{tabular}{|c|c||c|c|c||c|c|c|c|c|} 
                
 \hline 
$\delta$ &$T_c$& $\rho_0$&$R_{H0}$ &$10^4B$ & $10^4\chi_0$ & $T^*$& $\omega$  & $T_1$   \\ 
& K & $m\Omega cm$&$\frac{10^{-9}m^3}{C}$& $\frac{emu}{mole}$ & 
$\frac{emu}{mole}$ &K&K  &K 

\\ \hline
  0.05 &  90.7 & 1.8  & 0.45&&& 144&447 & 332   
\\ \hline
  0.12 &   93.7 &&&  2.6 &2.1&155 &&
\\ \hline
  0.19 &  87 &3.4 & 0.63 &4.5&1.6& 180&477& 454 
\\ \hline
  0.23 &  80.6& 5.7 & 0.74&&&210& 525  & 586 
\\ \hline
  0.26 &   78 & & &5.4 &1.5&259&&    
\\ \hline
  0.28 & 68.6 & 8.9& 0.81&&&259& 594  & 786
\\ \hline
  0.38 & 61.9 && &7.2 &1.4& 348&& 
\\ \hline
  0.39 & 58.1& 17.8& 0.96&&&344& 747 & 1088 
\\ \hline
  0.51 & 55 && & 9.1 &1.3& 494&& 
\\ \hline
\end{tabular} 

\vspace{01.5mm}

\begin{figure}
\begin{center}
\includegraphics[angle=-0,width=0.47\textwidth]{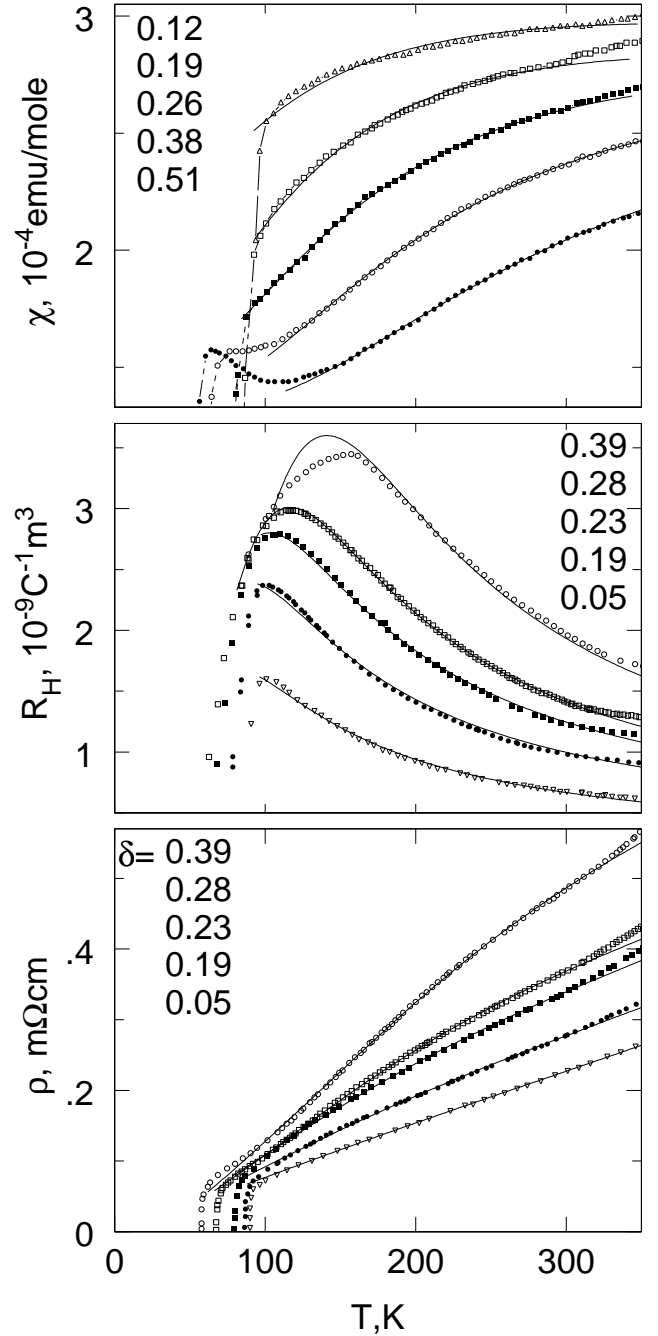}
\vskip -0.5mm
\caption{Uniform magnetic susceptibility, $\chi(T)$, Hall ratio, $R_H(T)$ and resistivity, $\rho(T)$, of underdoped $YBa_{2}Cu_{3}O_{7-%
\delta }$ fitted by the theory; see Table for the parameters. }
\end{center}
\end{figure}

As shown in Fig.1 and Fig.2, the model describes remarkably
well the experimental data with the parameters in this range (Table), in
particular the unusual maximum of the Hall ratio, well above $T_{c}$ (Fig.1),
and the non-linear temperature dependence of the in-plane resistivity. The
maximum of $R_{H}(T)$ is due to the contribution of thermally excited polarons
into transport, and the temperature dependence of the in-plane resistivity
below $T^{\ast }$ is due to this contribution and the combination of the
carrier-carrier and carrier-phonon scattering.  It is also quite remarkable that the
characteristic phonon frequency from the resistivity fit (Table) decreases
with doping as observed in the neutron scattering experiments 
and the pseudogap $T^{\ast }$ shows the doping behaviour as observed in
other experiments \cite{mih}. The temperature dependences of $R_{H}(T),$ $%
\rho (T)$ and $\chi (T)$ in underdoped $La_{2-x}Sr_{x}CuO_{4}$ and in other
underdoped cuprates are very similar to $YBa_{2}Cu_{3}O_{7-\delta }$, when
the temperature is rescaled by the pseudogap. It should be noted that adding a triplet bipolaron
band \cite{MOTT,mic} could improve the fit further (in particular, above room temperature) however increasing  the
number of fitting parameters.

To conclude: we applied the multi-polaron approach based on the
extension of the BCS theory to the strong-coupling regime \cite{SANF} to
describe peculiar normal state kinetics of underdoped cuprates. The low
energy physics in this regime is that of small bipolarons and thermally
excited polarons. Using this approach, we have explained the temperature
dependence of the Hall ratio, the in-plane resistivity and the bulk magnetic
susceptibility of underdoped cuprates. A direct measurement of the double
elementary charge $2e$ on carriers in the normal state could be decisive. In
1993, Mott and Alexandrov \cite{NEV} discussed the thermal conductivity $%
\kappa $; the contribution from the carriers provided by the Wiedemann-Franz
ratio depends strongly on the elementary charge as $\sim(e^{\ast })^{-2}$
and should be significantly suppressed in the case of $e^{\ast}=2e$.
Recently, a new way to determine the Lorenz number has been applied by Zhang
et al. \cite{zha}, based on the thermal Hall conductivity. As a result, the
Lorenz number has been directly measured in $YBa_{2}Cu_{3}O_{6.95}$.
Remarkably, the measured value of $L$ just above $T_{c}$ is the same as
predicted by the bipolaron model, $L\approx 0.15L_{e}$ ( $L_{e}$ is the
conventional Lorenz number). A breakdown of the Wiedemann-Franz law has been
also explained in the framework of the bipolaron model \cite{lor}.

This work was supported by the Leverhulme Trust (grant F/00261/H) and by the
Royal Society (grant ref: 15042). We are grateful to J.R.~Cooper for
helpful discussion of his experiments, and to A.J. Leggett for elucidating his theorem.

\end{document}